\documentclass[10pt,twocolumn,showpacs,amsmath,amssymb,osajnl,floatfix,superscriptaddress]{revtex4-1}

\usepackage{graphicx}
\usepackage{dcolumn}
\usepackage{bm}
\usepackage{color}
\usepackage{txfonts}
\usepackage{microtype}

\begin{document}

\author{Yan-Fei Li}	\affiliation{MOE Key Laboratory for Nonequilibrium Synthesis and Modulation of Condensed Matter, School of Science, Xi'an Jiaotong University, Xi'an 710049, China}
\author{Ren-Tong Guo}	\affiliation{MOE Key Laboratory for Nonequilibrium Synthesis and Modulation of Condensed Matter, School of Science, Xi'an Jiaotong University, Xi'an 710049, China}
\author{Rashid Shaisultanov}
\affiliation{Max-Planck-Institut f\"{u}r Kernphysik, Saupfercheckweg 1,
	69117 Heidelberg, Germany}
\author{Karen Z. Hatsagortsyan}
\affiliation{Max-Planck-Institut f\"{u}r Kernphysik, Saupfercheckweg 1,	69117 Heidelberg, Germany}
\author{Jian-Xing Li}\email{jianxing@xjtu.edu.cn}
\affiliation{MOE Key Laboratory for Nonequilibrium Synthesis and Modulation of Condensed Matter, School of Science, Xi'an Jiaotong University, Xi'an 710049, China}

\bibliographystyle{apsrev4-1}

\title{Single-shot determination of spin-polarization for ultrarelativistic electron beams\\
	 via nonlinear Compton scattering}

\date{\today}

\begin{abstract}

Impacts of spin-polarization of an ultrarelativistic electron beam head-on colliding with a strong  laser pulse on emitted photon spectra and electron dynamics have been investigated in the quantum radiation regime. We simulate photon emissions quantum mechanically and electron dynamics semiclassically via taking spin-resolved radiation probabilities in the local constant field approximation. A small ellipticity of the laser field brings about an asymmetry in
angle-resolved photon spectrum, which sensitively relies on the polarization of the electron beam.
The asymmetry is particularly significant in high-energy photon spectra, and is employed for the  polarization detection  of a high-energy electron beam with extraordinary precision,
e.g., better than 0.3\%   for a few-GeV electron beam at a density of the scale of $10^{16}$ cm$^{-3}$ with currently available strong laser fields.
This method  demonstrates a way of single-shot determination of polarization for ultrarelativistic electron beams via nonlinear Compton scattering. A similar method based on the asymmetry in the electron momentum distribution after the interaction due to  spin-dependent radiation reaction  is proposed as well.

\end{abstract}

\maketitle

Relativistic spin-polarized electron beams are extensively employed in nuclear physics and high-energy physics, e.g., to determine the neutron spin structure \cite{Anthony_1993}, to probe nuclear structures \cite{Abe_1995}, 
to generate polarized photons 
and positrons \cite{Maximon_1959},
to study  parity violation \cite{Androic_2018},
and to explore new physics beyond the Standard Model \cite{Moortgat2008}. They are generally produced  either via  an indirect method, first extracting polarized electrons from a photocathode \cite{Pierce_1976} or spin filters \cite{Batelann_1999,Dellweg_2017, Dellweg_2017PRA}
and then accelerating them, e.g., via laser wakefield acceleration \cite{Wen_2018},
or via a direct way, transversely polarizating
a relativistic electron beam in a storage ring via radiative polarization (Sokolov-Ternov effect) \cite{Sokolov_1964,Sokolov_1968,Baier_1967,Baier_1972,Derbenev_1973}.
Relativistic polarized positrons are commonly generated by Compton scattering or Bremsstrahlung of circularly polarized lasers and successive pair creation \cite{Omori_2006,Abbott_2016}. And, spin rotation systems can be utilized to alter the polarization direction 
 \cite{Buon_1986}.

With rapid developments of strong laser techniques,  stable (energy fluctuation 
$\sim$ 1\%) ultraintense (peak intensity $\sim 10^{22}$ W/cm$^2$, and magnetic field  $\sim 4\times 10^5$ Tesla) ultrashort (duration $\sim$ tens of fs) laser pulses have been generated \cite{Vulcan,ELI,Exawatt,Yanovsky2008}. Spin effects in nonlinear Compton scattering in such strong laser fields have been widely studied \cite{Panek_2002,Kotkin_2003,Karlovets_2011,Boca_2012,Krajewska_2013}. Due to the symmetry of the laser fields the radiative polarization of an electron beam, similar to the Sokolov-Ternov effect, is vanishing in a mononchromatic laser field \cite{Ivanov_2004} and rather small in the laser pulse \cite{Seipt_2018}, but can be extremely large in a model setup of strong rotating  electric fields \cite{Sorbo_2017,Sorbo_2018}.
Recently, the feasibility of significant polarization of an electron beam  in  currently achievable elliptically polarized (EP) laser pulses has been demonstrated due to spin-dependent radiation reaction (the polarization  higher than 70\% can be reached) \cite{Li_2018spin}.
The positrons from pair production  can also be highly polarized   in a similar setup (polarization up to 90\%) \cite{Wan_2019}, or in an asymmetric two-color laser field \cite{Chen_2019}.

The experiments with polarized electrons require a high precision and reliable polarimetry. Currently, the polarimetry for relativistic electron beams employs the following physical principles: Mott scattering \cite{Mott_1929}, M{\o}ller scattering \cite{Moller_1932, Cooper_1975}, linear Compton scattering  \cite{Klein_1929}, and synchrotron radiation \cite{Barber1993,Belomesthnykh_1984}. The polarization  of relativistic electrons is detected via  asymmetries in electron or photon momentum distribution.
However, the Mott and M{\o}ller polarimeters are only applicable  at low  energies ($< 10$ MeV) \cite{Gay1992,Tioukine2011} and at low currents ($\lesssim 100 \,\mu$A, due to target heating and subsequent depolarization at higher beam currents)  \cite{Gaskell2007,Hauger20013,Aulenbacher2018}, respectively,  and, the Compton polarimeter usually has to collect a 
large amount ($\gtrsim 10^5$ \cite{Beckmann_2002,Escoffier2005,Narayan_2016}) of laser shots to reach a small statistical uncertainty of $\sim 1\%$, due to low electron-photon collision luminosity.  
For low-repetition-rate dense ultrarelativistic electron beams, e.g., produced via strong laser pulses \cite{Li_2018spin, Wen_2018, Wan_2019, Chen_2019} with an energy $\sim$ GeV, a total charge $\sim$ pC, a duration $\sim 10$ fs (current $\gtrsim 100$ A), and a repetition rate ($\sim$ Hz), those first three methods are  inapplicable. In addition, their precisions for few-GeV electron beams are typically worse than 0.5\% \cite{Abe_2000, Friend_2012,Narayan_2016,Hauger_2001}, which cannot satisfy stringent request of proposed high-energy experiments, e.g.,  $\lesssim 0.4$\% \cite{Moller_2014axiv, Narayan_2016}. 
The polarimetry with synchrotron radiation is relatively slow (measurement time $\sim 1$ s) and not very accurate (precision $\sim 4\%$) \cite{Belomesthnykh_1984}, and finally can be carried out only in a large-scale synchrotron facility. Thus, a more efficient polarimetry with a  better precision and applicable for low-repetition-rate ultrarelativistic electron beams is still a challenge.

\begin{figure}[t]	
\setlength{\abovecaptionskip}{-0.0cm}  	
\includegraphics[width=1\linewidth]{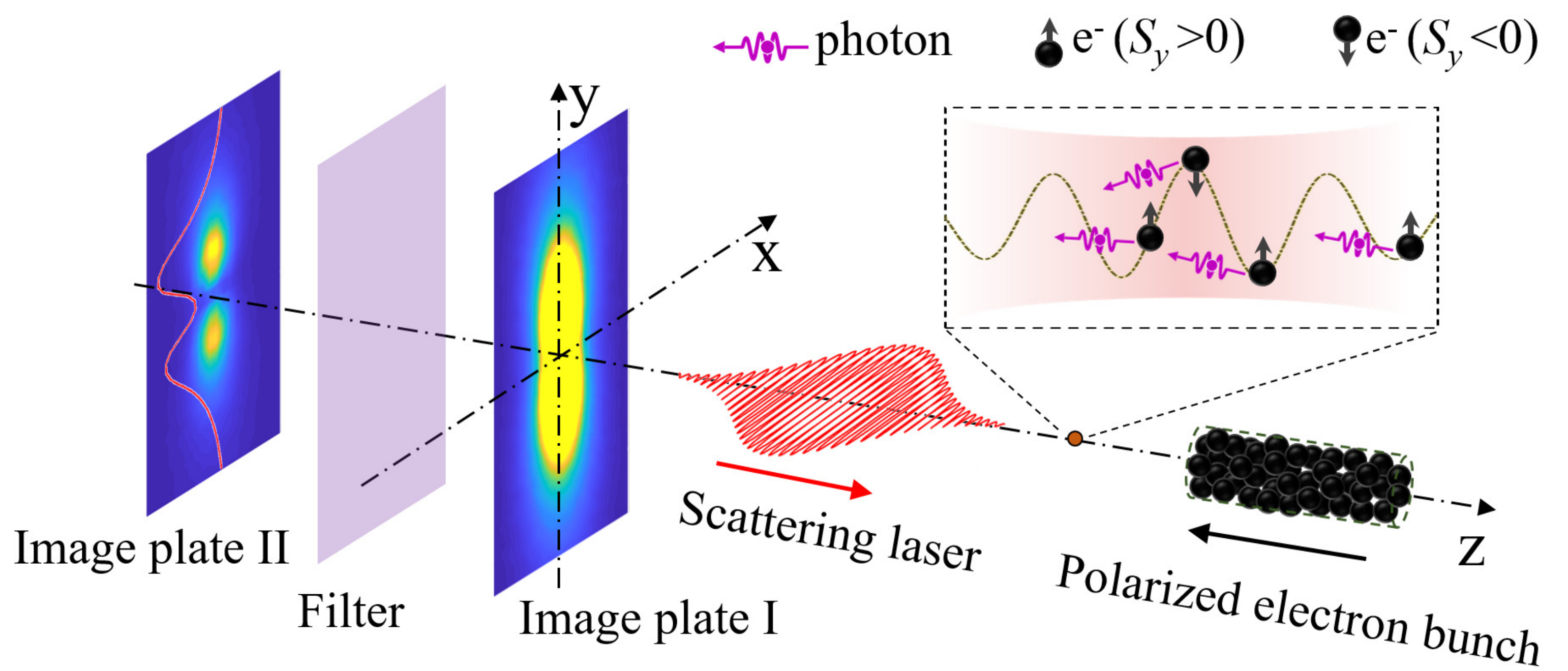}
	\caption{ Scenario of a  polarimetry of nonlinear Compton scattering. A strong EP laser pulse, propagating along $+z$ direction, collides with a transversely polarized (in $y$ axis) electron bunch. The major axis of the polarization ellipse is along $x$-axis.  The black curve in the sub-panel indicates the transverse laser field. Asymmetric angle-resolved spectra of all photons and filtered high-energy photons are shown on Image plate I and Image plate II, respectively. }
	\label{fig1} 
\end{figure}

In ultraintense laser fields, the Compton scattering  would reach the nonlinear realm due to multiphoton absorption \cite{Goldman_1964, Nikishov_1964, Ritus_1985}, which offers new paths for polarimetry, being especially attractive because of the remarkable increase of the number of emitted photons \cite{Piazza2012}.

In this Letter, we theoretically develop a new polarimetry method based on  
nonlinear Compton scattering, which can determine the polarization of a dense ultrarelativistic  electron beam via a single-shot interaction with a strong laser pulse, reaching a precision better than 0.3\%, see the interaction scenario in Fig.~\ref{fig1}. 
We consider an EP laser pulse  of currently available intensity head-on colliding with a polarized ultrarelativistic electron bunch in the quantum radiation regime. Because of  the spin-dependence of radiation probabilities, the electron most probably emits photons when its spin is anti-parallel to the laser magnetic field in its rest frame (chosen as the spin quantization axis (SQA)). Consequently, the photon emission intensity of the polarized electron beam in adjacent half laser cycles is asymmetric. Moreover, in this specific EP setup the photons from different half cycles are emitted in opposite directions with respect to the minor polarization axis ($y$-axis) of the laser field, creating asymmetric angular distribution of radiation, see detailed explanation below in Fig.~\ref{fig3}.
Since radiative spin effects are particularly conspicuous for high-energy photons, after filtering a more visible asymmetric spectrum of high-energy photons shows up, which is exploited for  the polarization determination.\\

We employ a Monte Carlo method to simulate
photon emissions during the electron semiclassical dynamics in external laser field \cite{Li_2018spin}, which is  based on the spin-resolved quantum radiation probabilities in the local constant field approximation (LCFA), valid at the invariant laser field parameter $\xi\equiv |e|E_0/(m\omega_0 c)\gg 1$   \cite{Ritus_1985, Baier1998}, where $E_0$ and $\omega_0$ are the amplitude and frequency of the laser field, respectively, $e$ and $m$ the electron charge and mass, respectively, and $c$ is the speed of the light in vacuum. The employed radiation  probabilities in LCFA  
are derived with the QED operator method of Baier-Katkov \cite{Baier_1973} and depend on the electron spin  vectors before and after radiation, ${\bf S}_{i}$ and ${\bf S}_f$ ($|{\bf S}_{i,f}|=1$) \cite{supplemental}. 
Summing over ${\bf S}_f$,  the radiation probability  depending on the initial spin is obtained (summed up by photon polarization) \cite{Baier_1973}:
	\begin{eqnarray}\label{Wspin2}
	\frac{{\rm d^2}{\overline{W}}_{fi}}{{\rm d}u{\rm d}\eta}&=&8W_R\left\{-(1+u){\rm IntK}_{\frac{1}{3}}(u')
	+(2+2u+u^2){\rm K}_{\frac{2}{3}}(u')\right.\nonumber\\
	&&\left.-u{\bf S}_i\cdot\left[{\bm\beta}\times\hat{{\bf a}}\right]{\rm K}_{\frac{1}{3}}(u')\right\},
	\end{eqnarray}
where, $W_R={\alpha m c}/\left[{8\sqrt{3}\pi\lambdabar_c\left( k\cdot p_i\right)}{\left(1+u\right)^3}\right]$, $u'=2u/3\chi$, $u= \omega_{\gamma}/\left(\varepsilon_i- \omega_{\gamma}\right)$, ${\rm IntK}_{\frac{1}{3}}(u')\equiv \int_{u'}^{\infty} {\rm d}z {\rm K}_{\frac{1}{3}}(z)$,  ${\rm K}_n$ is the $n$-order modified Bessel function of the second kind,  $\alpha$  the fine structure constant, $\lambdabar_c=\hbar/mc$ the Compton wavelength,  $\omega_{\gamma}$ the emitted photon energy, $\varepsilon_i$ the electron energy before radiation,   $\eta=k\cdot r$ the laser phase, ${\bm \beta}$ the electron velocity normalized by $c$, $p_i$, $r$, and $k$  are  the four-vectors of the electron momentum  before radiation, coordinate, and laser wave-vector, respectively, and $\hat{{\bf a}}={\bf a}/|{\bf a}|$ with the acceleration ${\bf a}$. The case when ${\bf S}_{i}$ and ${\bf S}_f$ are along the magnetic field in the rest frame of electron is given in \cite{King_2015}. 
Averaging Eq.~(\ref{Wspin2}) by ${\bf S}_{i}$, the widely used radiation probability is obtained \cite{Sokolov2010,Elkina2011,Ridgers_2014,Green2015,Harvey_2015}. 
The probabilities are characterized by the quantum parameter  $\chi\equiv |e|\hbar\sqrt{(F_{\mu\nu}p^{\nu})^2}/m^3c^4$ \cite{Ritus_1985}, where  $F_{\mu\nu}$ is the field tensor, and $\hbar$  the Planck constant.  When the electron counterpropagates with the laser beam, $\chi\approx 2(\hbar\omega_0/mc^2)\xi\gamma$, with the electron Lorentz factor $\gamma$. The spin dynamics due to photon emissions are described in the  quantum jump approach \cite{Molmer_1996,Plenio_1998}. 
After a photon emission, the electron spin state is stochastically collapsed into one of its basis states defined with respect to the instantaneous  SQA (along ${\bm\beta}\times\hat{{\bf a}}$) \cite{supplemental}. Between photon emissions, the electron dynamics in the external laser field is described by Lorentz equations, and the spin precession is governed by the Thomas-Bargmann-Michel-Telegdi  equation
\cite{Thomas_1926,Thomas_1927,Bargmann_1959, Walser_2002, supplemental}. \\

\begin{figure}[t]
		\setlength{\abovecaptionskip}{-0.0cm} 
	\includegraphics[width=1.0\linewidth]{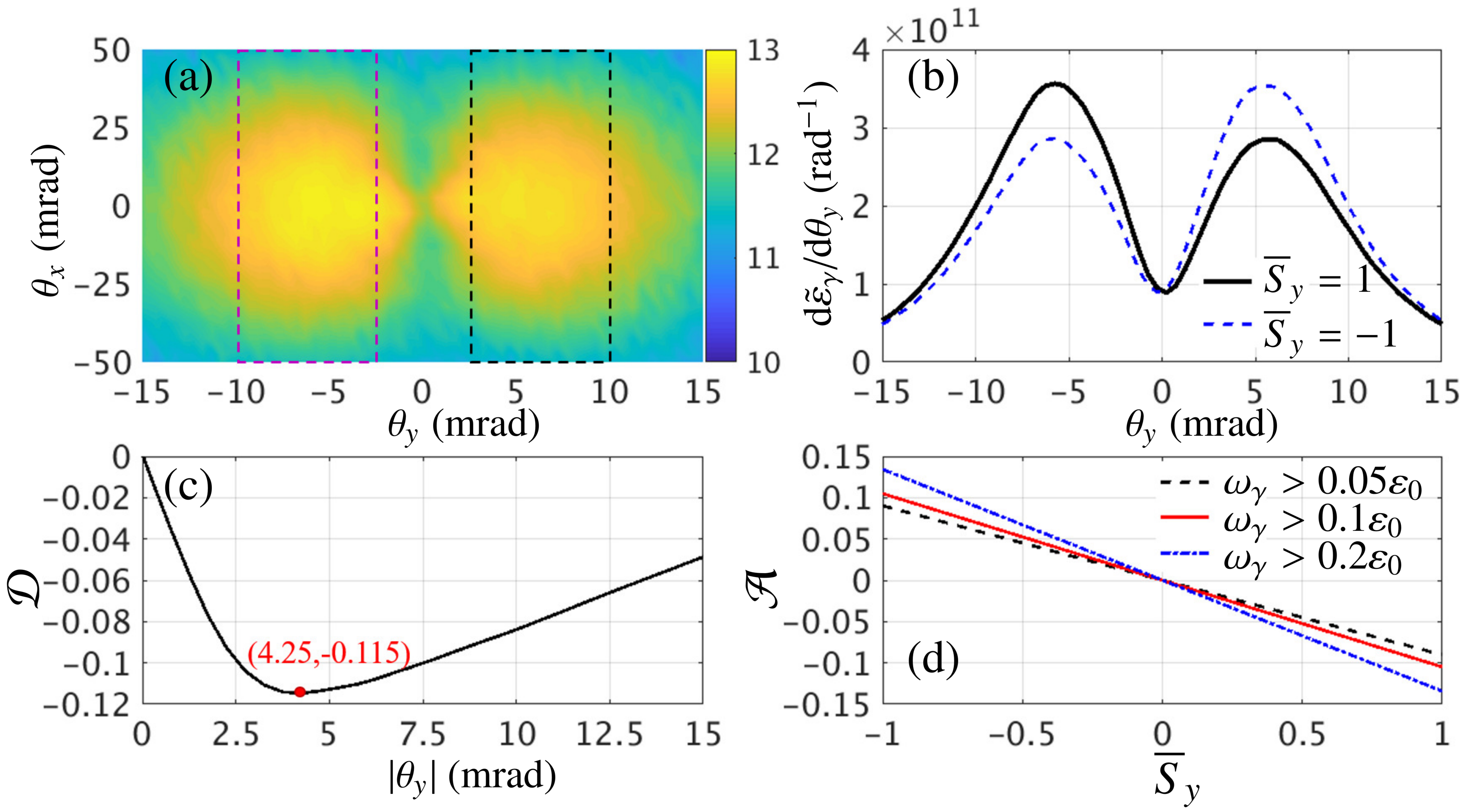}
	\caption{(a) Angle-resolved spectrum of selected high-energy photons ($\omega_\gamma > 0.1 \varepsilon_0$): $\log_{10}[d^2\varepsilon_\gamma/(mc^2d\theta_xd\theta_y)$] (rad$^{-2}$) vs. the transverse deflection angles $\theta_x={\rm arctan}(p_x/p_z)$ and $\theta_y={\rm arctan}(p_y/p_z)$,
	the initial average polarization $\overline{S}_y=1$.   
		(b)  d$\tilde{\varepsilon}_\gamma/$d$\theta_y=\int{\rm d}^2\varepsilon_\gamma/({\rm d}\theta_x{\rm d}\theta_y) {\rm d}\theta_x$ vs. $\theta_y$, with $\overline{S}_y=1$ (black-solid) and $\overline{S}_y=-1$ (blue-dashed), respectively. (c) Differential asymmetry $ \mathcal{D}$ for the case of $\overline{S}_y=1$. (d) Asymmetry  $\mathcal{A}$ vs $\overline{S}_y$. The photons used to calculate $\mathcal{A}$ are shown in the boxes in (a). The laser and electron beam parameters are given in the text. }
	\label{fig2} 
\end{figure}

Polarization determination via nonlinear Compton scattering is illustrated  in Fig.~\ref{fig2}.  
We employ a tightly-focused EP laser pulse with a Gaussian temporal profile. 
The spatial distribution of the electromagnetic fields takes into account up to $(w_0/z_r)^3$-order of the nonparaxial solution \cite{Salamin2002, supplemental}, where $w_0$ is the laser beam waist size, and $z_r$ the Rayleigh length.
The laser peak intensity  $I_0\approx1.38\times10^{22}$ W/cm$^2$ ($\xi=100$), wavelength $\lambda_0=1$ $\mu$m, pulse duration $\tau = 5T_0$,  $w_0=1.5$ $\mu$m, and  ellipticity $\epsilon=|E_y|/|E_x|=0.25$. The feasibility of such elliptical  polarization of ultrastrong laser  beams is  demonstrated in \cite{Gonzalez-Izquierdo_2016, Rodrigues_2019}.
The cylindrical electron bunch propagates at a polar angle $\theta_e=180^{\circ}$ (with respect to the laser propagation direction) and an azimuthal angle $\phi_e=0^{\circ}$ with an angular divergence of 1 mrad. The electron initial kinetic energy is $\varepsilon_0=1$ GeV, and the energy spread $\Delta \varepsilon_0/\varepsilon_0 =0.06$. In these conditions $\chi_{\rm max}\approx 0.4$, and the pair production is neglected.   The bunch radius  $w_e= 2\lambda_0$, length $L_e = 15\lambda_0$, and density $n_e\approx 5.3\times10^{16}$ cm$^{-3}$
with a transversely Gaussian and longitudinally uniform distribution, which
can be obtained by  laser wakefield accelerators \cite{Leemans2014,Leemans_2019} and polarized via radiative spin effects \cite{Li_2018spin, Wan_2019, Chen_2019}. 
Note that the case of relatively low-density electron bunchs produced by conventional accelerators or storage rings is also applicable \cite{supplemental}.

Figure~\ref{fig2}(a) demonstrates an asymmetric angle-resolved spectrum for high-energy photons $\omega_\gamma>0.1\varepsilon_0$.
The asymmetry is more visible in the spectrum  integrated over $\theta_x$, see Fig.~\ref{fig2}(b). The spectra for $\overline{S}_y=\pm 1$ cases are the most asymmetric, and other cases of $-1<\overline{S}_y<1$ would appear between them. 
For the quantitative characterization of asymmetry we introduce the differential asymmetry parameter $ \mathcal{D}=($d$\tilde{\varepsilon}_\gamma^{+}/$d$\theta_y-$d$\tilde{\varepsilon}_\gamma^{-}/$d$\theta_y)/($d$\tilde{\varepsilon}_\gamma^{+}/$d$\theta_y+$d$\tilde{\varepsilon}_\gamma^{-}/$d$\theta_y)$, between the values of d$\tilde{\varepsilon}_\gamma/$d$\theta_y$ at $\theta_y >0$ (``$+$'') and $\theta_y <0$ (``$-$'') with a same $|\theta_y|$, respectively, and the asymmetry parameter $\mathcal{A}=(\widetilde{\mathcal{E_\gamma}}^+-\widetilde{\mathcal{E_\gamma}}^-)/(\widetilde{\mathcal{E_\gamma}}^++\widetilde{\mathcal{E_\gamma}}^-)$, with $\widetilde{\mathcal{E_\gamma}}^{+}=\int_{0.0025}^{0.01}$(d$\tilde{\varepsilon}_\gamma/$d$\theta_y$)d$\theta_y$ and $\widetilde{\mathcal{E_\gamma}}^{-}=\int_{-0.01}^{-0.0025}$(d$\tilde{\varepsilon}_\gamma/$d$\theta_y$)d$\theta_y$. 
The differential asymmetry $\mathcal{D}$ is shown in Fig.~\ref{fig2}(c). As $|\theta_y|$ rises from 0 to 15 mrad, $|\mathcal{D}|$ first increases rapidly, reaches the peak of about 11.5\% at $|\theta_y|=$ 4.25 mrad, and then decreases slowly  to 5\%. For the asymmetry $\mathcal{A}$ the photons are selected in the regions of $-10$ mrad $\leq\theta_y\leq-2.5$ mrad and $2.5$ mrad $\leq\theta_y\leq 10$ mrad, where  $\mathcal{D}$ is apparently large. This angular region exceeds the uncertainty angle of the electron beam $\theta_{uncert}\sim 1/\gamma\approx0.7$ mrad \cite{supplemental}, as well as the currently achievable angular resolution for gamma-ray detection ($<1$ mrad) \cite{Cipiccia_2011,Phuoc_2012}.

The asymmetry parameter $\mathcal{A}$ is well suited to determine the polarization, see Fig.~\ref{fig2}(d).  
As $\overline{S}_y$ continuously increases from -1 to 1, $\mathcal{A}$ monotonously decreases  from 0.103 to -0.103 for the case of $\omega_\gamma > 0.1\varepsilon_0$. 
As the chosen photon energy decreases (increases) to $\omega_\gamma > 0.05\varepsilon_0$ ($0.2\varepsilon_0$), the slope of $\mathcal{A}$ curve decreases (increases) as well, with $\mathcal{A}_{\rm max}=0.09$ (0.134), which surpasses the asymmetry of
the Compton polarimeter ($<0.05$) \cite{Beckmann_2002,Escoffier2005,Narayan_2016}.
The precision of the polarization measurement  can be estimated via the statistical uncertainty  $\frac{\delta \mathcal{A}}{\mathcal{A}}\approx\frac{1}{\mathcal{A} \sqrt{N_\gamma}}$  
\cite{Placidi_1989}, reaching about 0.265\%, 0.31\% and 0.372\% for the cases of $\omega_\gamma > 0.05 \varepsilon_0$, $0.1 \varepsilon_0$ and $0.2 \varepsilon_0$, respectively, with the gamma-photon number $N_\gamma \approx$ $1.75 \times 10^7$, $1 \times 10^7$ and $0.4\times10^7$, respectively, in our simulations.   In the case of considering all photons, $\mathcal{A}_{\rm max}=0.078$, and the precision is about 0.16\% with $N_\gamma\approx 6.87\times10^7$ \cite{supplemental}. However, the experimental systematic uncertainty is inversely proportional to $\mathcal{A}$ \cite{Friend_2012}, which requires a large $\mathcal{A}$.

Note that the polarimetry can also be developed via using the asymmetry of the angular distribution of photon number, rather than photon energy in Fig.~\ref{fig2}, however, with a bit smaller $\mathcal{A}_{\rm max}$ \cite{supplemental}. \\

\begin{figure}
	\setlength{\abovecaptionskip}{-0.0cm} 
	\includegraphics[width=0.95\linewidth]{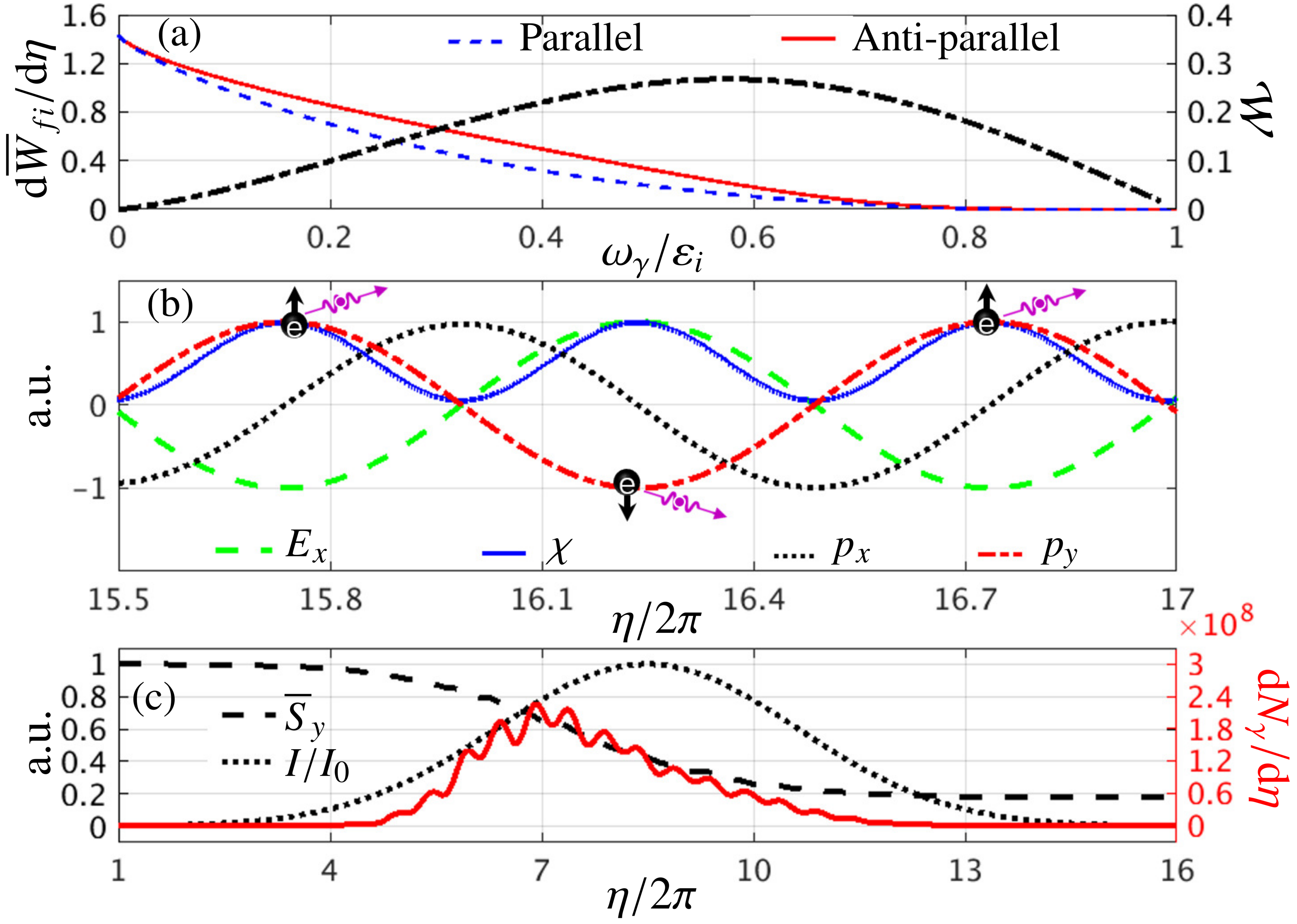}
	\caption{(a) Photon emission probabilities, ${\rm d} \overline{W}_{fi}/{\rm d}\eta$, with the spin vector parallel (blue-dashed) and anti-parallel (red-solid) to the instantaneous SQA, respectively, and their relative difference $\mathcal{W}\equiv ({\rm d} \overline{W}^{\rm anti}_{fi}/{\rm d}\eta-{\rm d} \overline{W}^{\rm paral}_{fi}/{\rm d}\eta)/({\rm d} \overline{W}^{\rm anti}_{fi}/{\rm d}\eta+{\rm d} \overline{W}^{\rm paral}_{fi}/{\rm d}\eta)$ (black-dash-dotted) vs.  $ \omega_{\gamma}/\varepsilon_i$. (b) Variations of $E_x$ (green-dashed),
 $\chi$ (blue-solid), and $p_x$ (black-dotted) and $p_y$ (red-dash-dotted) components of electron momenta, normalized by their maximal values in arbitrary units, with respect to $\eta$. (c) Normalized laser pulse intensity (black-dotted), $\overline{S}_y$ (black-dashed), and photon number density (red-solid) vs. $\eta$.  The laser and electron beam parameters in (a) and (c) are the same with those in Fig.~\ref{fig2} with $\omega_\gamma>0.1 \varepsilon_0$.  }
	\label{fig3} 
\end{figure}

The reasons of appearance of asymmetric spectra are analyzed in Fig.~\ref{fig3}. In Eq.~(\ref{Wspin2}), as the electron spin  ${\bf S}_{i}$ is anti-parallel to the instantaneous  SQA (along ${\bm\beta}\times\hat{{\bf a}}$), the photon emission probability is the largest and apparently larger than the parallel case,  see Fig.~\ref{fig3}(a). 
The relative difference of probabilities 
$\mathcal{W}$ is  remarkable for high-energy photons and reaches the peak, about 28\%, at $\omega_{\gamma}\approx0.6\varepsilon_i$. Thus, the asymmetry of high-energy photons is more visible, as indicated in Fig.~\ref{fig2}(d). However, the high-energy photons are much fewer, since the radiation probability declines gradually with the increase of $ \omega_{\gamma}$. Consequently, an appropriate photon energy $\omega_\gamma$ should be chosen, e.g., in Fig.~\ref{fig2}(d) the precision in  $\omega_\gamma>0.05  \varepsilon_i$ case is the best.

As demonstrated  in Fig.~\ref{fig3}(b), for the left-handed laser pulse $E_x$ has a $\pi/2$ phase delay with respect to $E_y$. The electron tranverse momentum in the laser field   ${\bf p}_\bot=-e{\bf A}(\eta)$, with the vector potential ${\bf A}(\eta)$, is ahead by $\pi/2$ with respect to the field ${\bf E}(\eta)$. Thus, compared with $E_x$ (green-dashed), $p_x$ (black-dotted) is ahead by $\pi/2$, and $p_y$ (red-dash-dotted) is ahead by $\pi$.  The radiation probability is determined by $\chi\propto\gamma \xi\propto E_x$. In the half cycles of $E_x>0$, 
the SQA is along ${\bm\beta}\times\hat{{\bf a}}\propto  e{\bm\beta}\times{\bf E}$,
 i.e., $+y$ direction, and consequently, the spin-down (with respect to $+y$ direction) electrons more probably emit photons, whose $p_y$  are certainly negative, but $p_x$ uncertain. In $E_x<0$, the SQA is along $-y$ direction, and the spin-up electrons more probably emit photons, whose $p_y$  are certainly positive.
 Thus, an asymmetric spectrum can appear along $y$ axis for the EP laser pulse, but not for the linearly polarized (LP) case. For the circularly polarized (CP) case, the SQA rotates, $p_y$ and $p_x$ components of emitted photons are both uncertain,  and consequently, the spectrum is symmetric in $x$-$y$ plane.

Due to radiative stochastic spin flips, the polarization of the electron beam is depressed during propagating through the laser pulse, see Fig.~\ref{fig3}(c), which could weaken the considered asymmetry. For ultrashort laser pulses, the chosen high-energy photons are mainly emitted at the front edge of the pulse, where the beam initial polarization is maintained well, and the asymmetric spectrum  corresponds to the initial polarization.  However, the asymmetric  spectrum of low-energy ($\omega_\gamma\ll \varepsilon_0$) photons, which are still substantially emitted at the back
edge of the pulse, can be significantly altered due to the beam depolarization.

Furthermore, we analyze the cases of larger energy spread $\Delta \varepsilon_0/\varepsilon_0=0.1$, larger angular divergence of 2 mrad, different collision angles $\theta_e=179^\circ$ and $\phi_e=90^\circ$, and 2\% fluctuation of laser intensity. All show stable and uniform results \cite{supplemental}. We underline that radiation reaction effects are not crucial for generating the asymmetry in photon spectra \cite{supplemental}. \\

 \begin{figure}[t]
 	\setlength{\abovecaptionskip}{-0.0cm}
 	\includegraphics[width=1.0\linewidth]{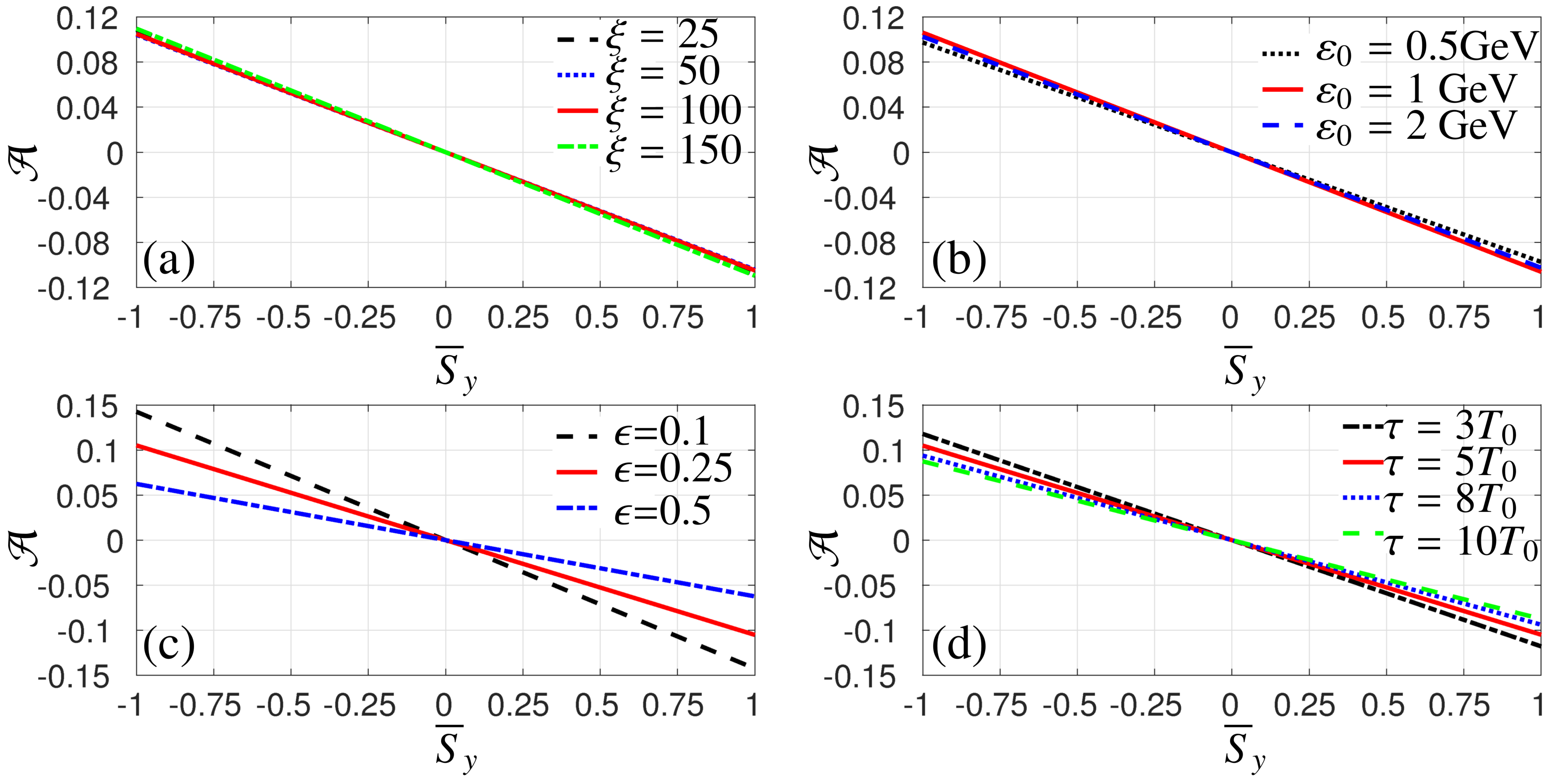}
 	\caption{(a)-(d): Impacts of $\xi$, $\varepsilon_0$, $\epsilon$ and $\tau$ on the asymmetry parameter $\mathcal{A}$. In (b), for the cases of $\varepsilon_0=0.5$ and 2 GeV, the detection $\theta_y$ ranges are $5<|\theta_y|<12.5$ and $1.25<|\theta_y|<8.75$, respectively. In (c), for the cases of $\epsilon=0.1$ and 0.5, $1<|\theta_y|<8.5$ and $5<|\theta_y|<12.5$, respectively. Other parameters are the same as those in Fig.~\ref{fig2}.}
 	\label{fig4} 
 \end{figure} 

\begin{figure}[t]
	\setlength{\abovecaptionskip}{-0.0cm}
	\includegraphics[width=1.0\linewidth]{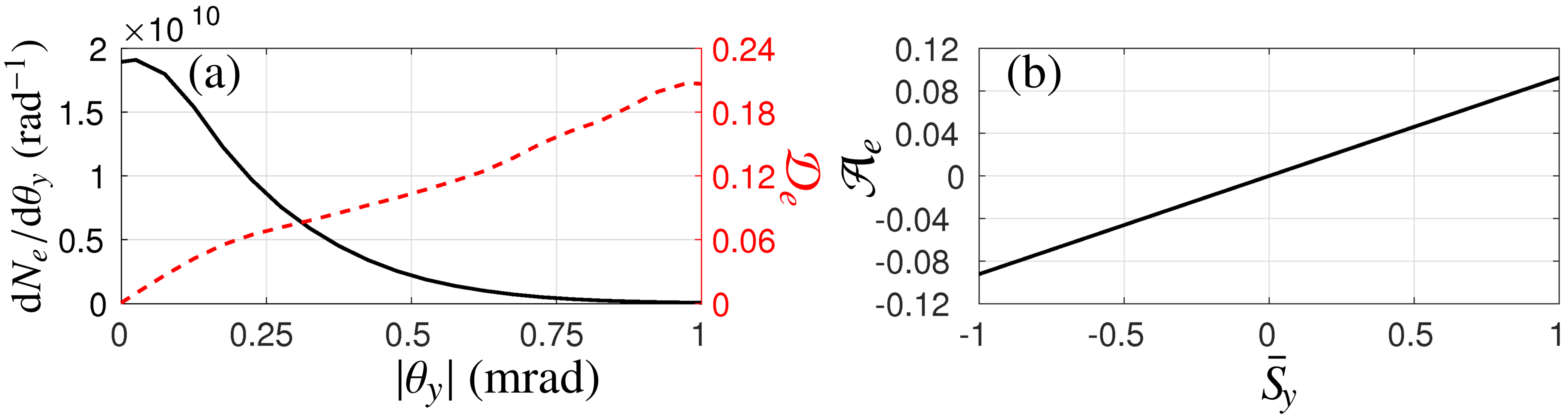}
	\caption{Polarization determination by the electron momentum distribution asymmetry. (a) d${N_e}/$d$\theta_y$  (black-solid) and $\mathcal{D}_e$ (red-dashed) vs. $\theta_y$ with $\overline{S}_y=1$. The calculation of $\mathcal{D}_e$ is the same as $\mathcal{D}$ in Fig.~\ref{fig2} except replacing $\varepsilon_\gamma$ by $N_e$. (b) The asymmetry parameter, $\mathcal{A}_e=(\widetilde{N_e}^+-\widetilde{N_e}^-)/(\widetilde{N_e}^++\widetilde{N_e}^-)$, with respect to  $\overline{S}_y$.  $\widetilde{N_e}^+=\int_{0.00025}^{0.001}$(d$N_e/$d$\theta_y$)d$\theta_y$, and $\widetilde{N_e}^{-}=\int_{-0.001}^{-0.00025}$(d$N_e/$d$\theta_y$)d$\theta_y$.  $\xi=20$, $\tau=20 T_0$, $\epsilon=0.1$, $w_0=2 \lambda_0$, $\varepsilon_0=4$ GeV, the electron beam divergence is 0.3 mrad, and other parameters are the same as those in Fig.~\ref{fig2}.}
	\label{fig5} 
\end{figure}

Impacts of the laser and electron beam parameters on the polarimetry 
is analyzed in Fig.~\ref{fig4}. First, as the laser intensity  rises, e.g., $\xi$ increases from 25 to 150 in Fig.~\ref{fig4}(a), not only the yield of high-energy photons increases, but the depolarization effect is also enhanced  \cite{supplemental}. Consequently, the $\mathcal{A}$ curve stays stable.  
The conditions of quantum regime $\chi\approx 2(\hbar\omega_0/mc^2)\xi\gamma\gtrsim 1$ and having a large amount of photons $N_\gamma\sim N_e\alpha\xi\tau/T_0\gg N_e$,  restrict the lower limit of $\xi$. $N_e$ is the electron number.  As $\xi\gg 1$ and $\chi\gg1$, the pair production and even cascade have to be considered, which would seriously affect high-energy photon spectra.      
As $\varepsilon_0$ increases, e.g., from 0.5 GeV to 2 GeV in Fig.~\ref{fig4}(b), the deflection angle of photons and the uncertainty angle of the electron beam both decrease. The corresponding detection angle ranges have to be adjusted to include the majority of high-energy photons and exceed $\theta_{uncert}$, and the $\mathcal{A}$ curve   changes slightly.
The photon deflection angle $\theta_y\sim p_y/p_z\propto E_y\propto\epsilon$. As $\epsilon$ rises, e.g., from 0.1 to 0.5 in Fig.~\ref{fig4}(c), $\theta_y$ rises as well, but the rotation effect of SQA in $x-y$ plane is enhanced (cf., the ultimate case of the CP laser), and consequently, the depolarization effect is enhanced as well. As $\epsilon$ is too small (cf., the ultimate case of the LP laser), the photons mix together, and the asymmetry is weakened and even removed. As the laser pulse duration increases, e.g., from $3 T_0$ to $10 T_0$ (currently achievable cases) in Fig.~\ref{fig4}(d), $\mathcal{A_{\rm max}}$ decreases, since the beam polarization declines  due to radiative spin effects, as demonstrated in Fig.~\ref{fig3}(c).

Finally, we point out that a similar method of polarimetry can be suggested via the electron momentum distribution asymmetry in the same EP laser setup, see Fig.~\ref{fig5}.
About 26\% electrons, which are in the range from 0.25 mrad to 1 mrad exceeding the uncertainty angle $\theta_{uncert}$, are used. $\mathcal{A}_{e, \rm max}\approx0.092$, and the precision is about 0.67\%. Compared with above method, this method gives a lower precision, because the considered electron number is smaller than the emitted photon number due to multiple photon emissions as $\xi\gg1$, but is more efficient for low laser intensities when $N_\gamma/N_e\sim \alpha\xi\tau/T_0\lesssim 1$. \\

In conclusion, we have developed a new method of polarimetry based on nonlinear Compton scattering in the quantum radiation regime.  The electron beam polarization can be measured via 
the angular asymmetry of high-energy gamma-photon spectrum in a  single-shot interaction of the electron beam with an EP strong laser pulse of currently achievable intensity. The precision for the polarization measurement is better than  0.3\% for dense few-GeV electron beams, evidently exceeding those of presently available common techniques.\\

 {\it Acknowledgement:}  This work is supported by the  Science Challenge Project of China (No. TZ2016099), the National Key Research and Development Program of China (Grant No. 2018YFA0404801), and the National Natural Science Foundation of China (Grants Nos. 11874295, 11804269). 
 
\bibliography{QEDspin}

\end{document}